\documentclass[superscriptaddress,onecolumn,secnumarabic,
amssymb,amsmath,nobibnotes,aps,prd,showkeys,showpacs,nofootinbib]{revtex4}

\usepackage[latin1]{inputenc}
\usepackage{graphicx}
\usepackage[english]{babel}

\usepackage{amsmath}
\usepackage{amssymb}
\usepackage{amsfonts}
\usepackage{colordvi}
\usepackage{psfrag}
\usepackage{color}

\begin{document}

\def\cL{{\cal L}}
\def\be{\begin{equation}}
\def\ee{\end{equation}}
\def\bea{\begin{eqnarray}}
\def\eea{\end{eqnarray}}
\def\beq{\begin{eqnarray}}
\def\eeq{\end{eqnarray}}
\def\tr{{\rm tr}\, }
\def\nn{\nonumber \\}
\def\e{{\rm e}}

\def\VEV#1{\left\langle #1\right\rangle}        


\def\bef{\begin{figure}}
\def\eef{\end{figure}}
\newcommand{\ans}{ansatz }
\newcommand{\eeqn}{\end{eqnarray}}
\newcommand{\bd}{\begin{displaymath}}
\newcommand{\ed}{\end{displaymath}}
\newcommand{\mat}[4]{\left(\begin{array}{cc}{#1}&{#2}\\{#3}&{#4}
\end{array}\right)}
\newcommand{\matr}[9]{\left(\begin{array}{ccc}{#1}&{#2}&{#3}\\
{#4}&{#5}&{#6}\\{#7}&{#8}&{#9}\end{array}\right)}
\newcommand{\matrr}[6]{\left(\begin{array}{cc}{#1}&{#2}\\
{#3}&{#4}\\{#5}&{#6}\end{array}\right)}
\newcommand{\cvb}[3]{#1^{#2}_{#3}}
\def\lsim{\raise0.3ex\hbox{$\;<$\kern-0.75em\raise-1.1ex
e\hbox{$\sim\;$}}}
\def\gsim{\raise0.3ex\hbox{$\;>$\kern-0.75em\raise-1.1ex
\hbox{$\sim\;$}}}
\def\abs#1{\left| #1\right|}
\def\simlt{\mathrel{\lower2.5pt\vbox{\lineskip=0pt\baselineskip=0pt
           \hbox{$<$}\hbox{$\sim$}}}}
\def\simgt{\mathrel{\lower2.5pt\vbox{\lineskip=0pt\baselineskip=0pt
           \hbox{$>$}\hbox{$\sim$}}}}
\def\unity{{\hbox{1\kern-.8mm l}}}
\newcommand{\eps}{\varepsilon}
\def\ep{\epsilon}
\def\ga{\gamma}
\def\Ga{\Gamma}
\def\om{\omega}
\def\omp{{\omega^\prime}}
\def\Om{\Omega}
\def\la{\lambda}
\def\La{\Lambda}
\def\al{\alpha}
\newcommand{\ov}{\overline}
\renewcommand{\to}{\rightarrow}
\renewcommand{\vec}[1]{\mathbf{#1}}
\newcommand{\vect}[1]{\mbox{\boldmath$#1$}}
\def\tm{{\widetilde{m}}}
\def\mcirc{{\stackrel{o}{m}}}
\newcommand{\Dm}{\Delta m}
\newcommand{\dm}{\varepsilon}
\newcommand{\tanb}{\tan\beta}
\newcommand{\nbar}{\tilde{n}}
\newcommand\PM[1]{\begin{pmatrix}#1\end{pmatrix}}
\newcommand{\up}{\uparrow}
\newcommand{\down}{\downarrow}
\def\omE{\omega_{\rm Ter}}
\def\caja{\mathsurround=0pt}
\def\eqalign#1{\,\vcenter{\openup2\jot \caja
        \ialign{\strut \hfil$\displaystyle{##}$&$
        \displaystyle{{}##}$\hfil\crcr#1\crcr}}\,}
%

\newcommand{\Dsusy}{{susy \hspace{-9.4pt} \slash}\;}
\newcommand{\DCP}{{CP \hspace{-7.4pt} \slash}\;}
\newcommand{\mc}{\mathcal}
\newcommand{\gr}{\mathbf}
\renewcommand{\to}{\rightarrow}
\newcommand{\gtc}{\mathfrak}
\newcommand{\wh}{\widehat}
\newcommand{\br}{\langle}
\newcommand{\kt}{\rangle}


\def\lsim{\mathrel{\mathop  {\hbox{\lower0.5ex\hbox{$\sim$}
\kern-0.8em\lower-0.7ex\hbox{$<$}}}}}
\def\gsim{\mathrel{\mathop  {\hbox{\lower0.5ex\hbox{$\sim$}
\kern-0.8em\lower-0.7ex\hbox{$>$}}}}}

\def\nn{\\  \nonumber}
\def\de{\partial}
\def\brf{{\mathbf f}}
\def\bbf{\bar{\bf f}}
\def\bF{{\bf F}}
\def\bbF{\bar{\bf F}}
\def\bA{{\mathbf A}}
\def\bB{{\mathbf B}}
\def\bG{{\mathbf G}}
\def\bI{{\mathbf I}}
\def\bM{{\mathbf M}}
\def\bY{{\mathbf Y}}
\def\bX{{\mathbf X}}
\def\bS{{\mathbf S}}
\def\bb{{\mathbf b}}
\def\bh{{\mathbf h}}
\def\bg{{\mathbf g}}
\def\bla{{\mathbf \la}}
\def\bmu{\mathbf m }
\def\by{{\mathbf y}}
\def\bmu{\mbox{\boldmath $\mu$} }
\def\bsig{\mbox{\boldmath $\sigma$} }
\def\bunity{{\mathbf 1}}
\def\cA{{\cal A}}
\def\cB{{\cal B}}
\def\cC{{\cal C}}
\def\cD{{\cal D}}
\def\cF{{\cal F}}
\def\cG{{\cal G}}
\def\cH{{\cal H}}
\def\cI{{\cal I}}
\def\cL{{\cal L}}
\def\cN{{\cal N}}
\def\cM{{\cal M}}
\def\cO{{\cal O}}
\def\cR{{\cal R}}
\def\cS{{\cal S}}
\def\cT{{\cal T}}
\def\eV{{\rm eV}}

\thispagestyle{empty}

{\hbox to\hsize{
\vbox{\noindent January 2017 \hfill IPMU17-0005 }}}

\noindent
\vskip2.0cm

\title{Energy conditions in Starobinsky supergravity}

\author{Andrea Addazi}

\affiliation{ Dipartimento di Fisica,
 Universit\`a di L'Aquila, 67010 Coppito AQ, Italy}
 
\affiliation{Laboratori Nazionali del Gran Sasso (INFN), 67010 Assergi AQ, Italy}

\author {Sergei V. Ketov}

\affiliation{Department of Physics, Tokyo Metropolitan University
Minami-ohsawa 1-1, Hachioji-shi, Tokyo 192-0397, Japan} 

\affiliation{Kavli Institute for the Physics and Mathematics of the Universe (Kavli IPMU), 
The University of Tokyo, Chiba 277-8568, Japan}

\affiliation{Institute of Physics and Technology, Tomsk Polytechnic University,
30 Lenin Ave., Tomsk 634050, Russian Federation}

\date{\today}

\begin{abstract}

We consider the classical energy conditions (strong, null and weak) in Starobinsky supergravity theories.
We study in detail the simplest Starobinsky supergravity model in the "old-minimal" supergravity setup 
and find  examples of violations of each energy condition for a certain range of the parameters due to extra 
scalar d.o.f. present in supergravity. We find the null and weak energy conditions to be preserved and the
strong energy condition to be violated in the physically relevant examples such as inflation and dark energy.

\end{abstract}
\pacs{04.65.+e,98.80.Cq}
\keywords{Modified Gravity, Supergravity, Energy Conditions, Inflation }

\maketitle

\section{Introduction}

In a metric theory of gravity, energy conditions lead to very deep implications 
on the fundamental issues such as causal space-time structure, space-time singularities, 
geodesic dynamics, gravitational attractiveness, etc. The energy conditions are also used in several important theorems of General Relativity (GR). For example, as is well known, Penrose's theorems  rely on some assumptions about the so-called Null Energy Conditions (NEC) \cite{P1,P2,P3}. 
A violation of the NEC may avoid singularities that are generically present in classical solutions to General Relativity equations of motion. On the other hand, a large violation of the NEC may also lead to dangerous consequences to the laws of BH thermodynamics, Cauchy's theorems, Cosmic Censorship,  wormholes, violations of causality and quantum space-time nucleation instabilities --- see, for example, 
\cite{Visser:1999de,Rubakov:2014jja} for reviews. 

Various energy conditions can be formulated in General Relativity  as bounds on the matter energy-momentum tensor, by contracting the latter with time-like
and null vectors. The mostly used are the so-called Strong Energy Condition (SEC), the Null Energy Condition (NEC) and the Weak Energy Condition (WEC).  Under general assumptions of GR,  those energy conditions are applied to Einstein's field equations and Raychaudhuri's equation (subject to Hawking-Ellis decomposition)  given by \cite{he} 
\begin{equation} \label{first}
\dot{\theta}+\frac{\theta^{2}}{3}+2(\sigma^{2}-\omega^{2})-\dot{W}^{a}_{;a}=-R_{ab}W^{a}W^{b}~~,
\end{equation}
where $2\sigma^{2}$ is the shear tensor, $\theta$ is the expansion scalar, $\omega^{2}$ is square of the vorticity tensor, $R_{ab}$ is Ricci tensor and $W_{a}$ is a time-like or a null-like 4-velocity vector. 

GR is an impressively successful theory of gravity, and it was verified against many experimental texts. However,  GR still has many unsolved problems as regards dark matter, dark energy, black holes, space-time singularities, cosmological inflation, etc. Though Einstein's field equations are rather complicated  PDE's, the Einstein-Hilbert action is obviously  the simplest consistent metric action leading to the  standard Newtonian limit. However, in many occasions as, for example, the early universe, the Newtonian limit is not
required, so that the Einstein-Hilbert action may be modified to a more general one under the conditions of locality
and general covariance.  One of the simple extensions is a replacement $R \rightarrow f(R)$, where $f$ is an analytical function of the Ricci scalar $R$, known as $f(R)$ gravity.  The $f(R)$-gravity leads to the higher derivative terms of the metric field in its equations of motion, unlike the minimal EH action that also has higher derivatives but
they disappear in the equations of motion. When demanding stability of $f(R)$ gravity, i.e. the absence of ghosts and tachyons, it is always possible to transform it into a scalar-tensor theory of gravity without higher derivatives  but with an extra scalar d.o.f. known as scalaron (see \cite{Nojiri:2006ri,Nojiri:2010wj,Capozziello:2011et,Ketov:2012yz} for reviews).

  The simplest example of a ghost- and tachyon-free $f(R)$ gravity is given by the famous Starobinsky model \cite{Starobinsky:1980te}
 with 
 \be \label{star}
S_{\rm Star.} = \int \mathrm{d}^4x\sqrt{-g} \left[ \frac{1}{2} R +\frac{1}{12M^2}R^2\right]~,
\ee
where we have used natural units with the reduced Planck mass $M_{\rm Pl}=1$. This model is
an excellent model of inflation, having very good agreement with the Planck data
\cite{Ade:2015lrj}. The Starobinsky model is geometrical (i.e. includes "gravity" only),
while its only (real) parameter $M$ can be identified with the scalaron mass.  When being considered as the
inflationary model, the mass parameter $M$ is fixed by the observational Cosmic 
Microwave Background (CMB) data as $M=(3.0 \times10^{-6})(\frac{50}{N_e})$ where $N_e$ is the e-foldings number.  The corresponding scalar potential of scalaron $\phi$ in the dual (scalar-tensor gravity) picture is given by
\begin{equation} \label{starp}
V(\phi) = \frac{3}{4} M^2\left( 1- e^{-\sqrt{\frac{2}{3}}\phi }\right)^2~.
\end{equation}
This scalar potential is bounded from below (actually, is also non-negative and stable), with its minimum corresponding to a Minkowski vacuum. The scalar potential (\ref{starp}) has a plateau of a positive height, which describes the inflationary era.
 
The energy conditions in $f(R)$-gravity have to be extended vs. Einstein's General Relativity 
\cite{Mimoso:2014ofa,Capozziello:2014bqa} because the scalaron field, propagating 
in $f(R)$-gravity, contributes to the effective energy-momentum tensor.  That is why, for instance, Capozziello {\it et al.} suggested to dub it as a {\it geometric fluid} in \cite{Capozziello:2014bqa}.

One may naively expect that the nice features of $f(R)$ gravity and Starobinsky model mentioned above would
be automatically valid in their supergravity extensions also, as well as the corresponding energy conditions.
We find in this paper that it is not the case. We confine ourselves to a supergravity extension of the Starobinsky
$(R+R^2)$-gravity only, and call it the Starobinsky supergravity. As was demonstrated in \cite{Ketov:2013dfa,Ferrara:2013pla}, the supergravity  extensions of $(R+R^{n})$-gravity with $n\geq 3$ are plagued with tachyons and are unstable. 
On the other hand, supergravity has gravitinos that can be good candidates for cold dark matter.~\footnote{A supergravity scenario in which a continuos spectrum of supermassive gravitinos is non-adiabatically produced during inflation, was considered in Refs.~\cite{Addazi:2016mtn,Addazi:2016bus}.}
We study SEC, NEC and WEC  in the Starobinsky supergravity and find constraints on their applicability. Unlike (modified) gravity theories, supergravity highly restricts possible couplings. Ghosts and tachyons can be avoided in the Starobinsky supergravity, and we always demand it.


\section{The Starobinsky supergravity}

The most general extension of the Starobinsky $(R+R^2)$-gravity  (\ref{star}) in curved N=1 superspace
of the old-minimal N=1 supergravity is given by \cite{Ketov:2013dfa}
\begin{equation}
\label{A}
S=\int d^{4}x d^{4}\Theta E^{-1}N(\mathcal{R},\bar{\mathcal{R}})+\left[\int d^{4}x d^{2}\Theta 2\mathcal{E}F(\mathcal{R})+h.c    \right]~,
\end{equation}
where the D-term is parameterized by a non-holomorphic real potential $N(\mathcal{R},\mathcal{R})$, and the 
F-term is parameterized by a holomorphic potential $F(\mathcal{R})$, in terms of the N=1 covariantly-chiral superfield 
$\mathcal{R}$ and its conjugate $\bar{\mathcal{R}}$ that contain the scalar curvature $R$ in their field components at
$\Theta^2$. 

A generic theory (\ref{A}) with arbitrary functions $N$ and $F$ can be transformed into the standard matter-coupled Einstein supergravity (without higher derivatives) with the matter given by two chiral superfields, the inflation superfield $\Phi$ and the
Goldstino supergield $S$ \cite{Cecotti:1987sa,Ketov:2013dfa}. It is the latter (dual formulation) that is usually considered in the literature. In the absence of a D-term, the theory (\ref{A}) is known as the chiral $F(\mathcal{R})$-supergravity \cite{Ketov:2010qz} that is dual to the matter-coupled Einstein supergravity with only one chiral matter (inflaton) superfield $\Phi$,  though the  chiral $F(\mathcal{R})$-supergravity cannot exactly reproduce the $(R+R^2)$ terms and the scalar potential (\ref{starp}).

In this paper we use the original formulation (\ref{A}) of the Starobinsky supergravity (corresponding to the Jordan frame), where its "pure supergravity" (or geometrical) nature is manifest, because it highly constrains the inflationary predictions.  The full bosonic action of the theory (\ref{A}), when all the fermionic terms vanish, was computed in Ref.~\cite{Ketov:2013dfa}.~\footnote{See also 
Refs.~\cite{Ferrara:1978rk,Kallosh:2013xya,Rocek:1978nb,Komargodski:2009rz,Antoniadis:2014oya}.}

To simplify our discussion and make it more transparent, we choose here the ansatz
\begin{equation}
 \label{ansatz}
N(\mathcal{R},\bar{\mathcal{R}})=\frac{12}{M^{2}}\bar{\mathcal{R}}\mathcal{R}-\frac{\zeta}{2}(\bar{\mathcal{R}}
\mathcal{R})^{2}~,\qquad  F(\mathcal{R})=f_{0}+3\beta \mathcal{R}~,
\end{equation}
with real positive parameters $M$, $\beta$ and $\zeta$, and a complex $f_{0}$, whose physical significance is explained below.
As regards the energy conditions, we believe that our ansatz  (\ref{ansatz}) captures the main features of a generic theory (\ref{A}) also.

The bosonic part of the supergravity action depends upon $(g_{mn},X,b_{a})$, where the complex scalar $X$ and the real axial vector $b_a$
represent the "auxiliary" fields of the standard (without higher derivatives) old-minimal N=1 supergravity, which become dynamical in the higher-derivative theory
(\ref{A}). More precisely, the chiral scalar $X$ is the superpartner (sGoldstino) of the Goldstino fermion arising due to supersymmetry breaking during inflation, whereas the divergence $\partial_ab^a$ of the "auxiliary" vector field in the action (\ref{A}) results in the presence of an additional propagating real scalar that complexifies scalaron (we call it sinflaton). Those four scalar d.o.f. exactly correspond to the spin-0 components of the two chiral N=1 superfields in the dual formulation of the theory as a matter-coupled Einstein supergravity. 
 
We demand the absence of ghosts and the existence of a (classical) ground state, i.e. the boundedness of the scalar potential from below.  
The bosonic part of our supergravity model reads 
\begin{equation}
 \label{model}
 \eqalign{
(-g)^{-1/2}\mathcal{L} & = \left(\frac{\beta}{2}-\frac{2}{M^{2}}|X|^{2}+\frac{11}{12}\zeta|X|^{4}\right)\left( R +\frac{2}{3}b_a^2\right)
+ \frac{4}{3}\left(  \frac{1}{12M^2}- \frac{\zeta}{72}\abs{X}^2\right)b_a^2R \cr
& + \left(  \frac{1}{12M^2}- \frac{\zeta}{72}\abs{X}^2\right)\left[ R^2 - (12\partial_a X)^2+ 4(D_ab^a)^2 +\frac{4}{9}(b_a^2)^2\right] \cr
&  + 48\left(  \frac{1}{12M^2}- \frac{\zeta}{144}\abs{X}^2\right)b^a{\rm Im}(X\partial_a X) - V(\bar{X},X)
}
\end{equation}
with $|X|^{2}=\bar{X}X$, and  the scalar potential 
\begin{equation}
 \label{xpoten}
V(\bar{X},X)=-6(f_{0}\bar{X}+\bar{f}_0X)   + 6\abs{X}^2\left(  -2\beta - \frac{8}{M^{2}}|X|^2+3\zeta(\abs{X}^{2})^2\right)~~.
\end{equation}

Scalaron arises after dualization of the $R^2$-term. As was demonstrated in Ref.~\cite{Ketov:2013dfa}, the divergence $\partial_ab^a$ obeys the generalized (by some non-linear terms)  Klein-Gordon equation with mass $M$, so that it represents the sinflaton indeed.  In what follows, we make a further simplification of ignoring sinflaton by setting $b_a=0$  because sinflaton enters the action 
via its space-time derivatives only, has the vanishing vacuum expectation values, $\VEV{\partial_ab^a}=\VEV{b^a}=0$, and does not change our final results.

Equation (\ref{model}) is obviously an extension of the Starobinsky model (\ref{star}), and one gets it precisely when $b_a=X=0$ and $\beta=1$. The role of the $\zeta$-term in eq.~(\ref{ansatz}) is also clear: $\zeta>0$ is
needed to stabilize the scalar potential (\ref{xpoten}) of $X$.~\footnote{The need of non-minimal terms in the dual version of
Starobinsky supergravity was observed in Ref.~\cite{Kallosh:2013xya}.} 
The vacuum expectation value $\VEV{X}\equiv X_0$ is determined by a
minimum of the scalar potential (\ref{xpoten}), which implies the equation $V_X=0$ that reads
\begin{equation} \label{crit}
\abs{X_0}^2\left(9\zeta\abs{X_0}^2 - \frac{16}{M^2}\right)X_0 - 2\beta X _0- f_0=0
\end{equation}
The presence of the linear term in (\ref{xpoten}), proportional to $f_{0}$, is compatible with our consistency conditions, since it does not destabilize the theory. The physical significance of $f_0\neq 0$ is breaking the R-symmetry. 
However, this term vanishes along the direction $f_{0}\bar{X}+\bar{f}_0X=0$ in the $X$-space. We find that this term does not affect our conclusions, so we set $f_0=0$ later in this Section, for simplicity. The vacuum solutions are 
$R$-symmetric in this case, under phase rotations of $X_0$.

Finally, the role of the parameter $\beta$ is to normalize the Einstein-Hilbert term in eq.~(\ref{model}) by demanding
\begin{equation}\label{betanorm}
\beta = 1 +\frac{4}{M^2}\abs{X_0}^2-\frac{11}{6}\zeta\abs{X_0}^4~~.
\end{equation}

Physical properties of the Starobinsky supergravity depend upon the scales, $X_0$, $M$ and $\zeta$. For instance, the scalaron mass $M_0$ is determined by the relation
\begin{equation} \label{smass}
 \frac{1}{12M^2}- \frac{\zeta}{72}\abs{X_0}^2= \frac{1}{12M_0^2}~~,
\end{equation}
while $M_0$ should be of the order $10^{-6}$, as is required by the CMB observations. It implies
\begin{equation} \label{bound}
\abs{X_0}^2 < \frac{6}{\zeta M^2}~~.
\end{equation}

In the case of $f_0=0$ the R-symmetry is preserved, the criticality condition (\ref{crit}) always has a solution $X_0=0$ that leads to $V(X_0)=0$ and $\beta=1$. However, this critical point
does not correspond to a minimum of the scalar potential, as is also clear from (\ref{xpoten}).~\footnote{The same conclusion arises with a small $f_0\neq 0$ and $X_0\approx -\frac{f_0}{2\beta}$.} Our model in this case appears to be very close to the original Starobinsky model (\ref{star}), but has a negative cosmological constant at the minimum of the scalar potential (\ref{xpoten}), i.e.  $V(X_0)<0$ because $\beta>0$ and our scalar potential is bounded from below.

The only way to avoid $V(X_0)<0$ is to have $\beta<0$, i.e. to go away from the "canonical" values, $\beta=1$ and $X_0=0$. Then (\ref{xpoten}) implies
\begin{equation}  \label{conm}
 \abs{X_0}^2\left[ 9\zeta   \abs{X_0}^2\ - \frac{16}{M^2} \right] = 2\beta <0~~.
\end{equation} 
It is a quadratic equation with the solution
\begin{equation}  \label{cons}
 \abs{X_0}^2= \frac{9}{8\zeta M^2}\left[ 1 + \sqrt{ 1 + \frac{9}{32}\beta\zeta M^4}\; \right]~~.
\end{equation} 
This solution implies 
\begin{equation}  \label{rest2}
 \beta \geq -\frac{32}{9\zeta M^4}
\end{equation} 

The corresponding value of the scalar potential is
\begin{equation}  \label{minval}
V(X_0) = 8 \abs{X_0}^2\left[ - \frac{2}{M^2} \abs{X_0}^2 -\beta \right]~~.
\end{equation} 
Hence, having $V(X_0)\geq 0$ also implies 
 $\beta \leq -2 \abs{X_0}^2/M^2$. Moreover, a consistency between (\ref{betanorm}) and (\ref{conm}) allows us to exclude $\beta$ and
get another quadratic equation for $\abs{X_0}^2$, whose solution is 
\begin{equation}  \label{veve2}
  \abs{X_0}^2= \frac{18}{19\zeta M^2}\left[ 1 + \sqrt{ 1 + \frac{19}{108}\zeta M^4}\; \right]~~.
\end{equation} 
Equations (\ref{bound}), (\ref{cons}) and (\ref{veve2}) together tightly constrain the allowed values of the parameters 
$\zeta M^4$ and $\beta$, though they exist within the bounds
\begin{equation}  \label{bound1}
0<\zeta M^4\leq \frac{108}{19}\left[ \left(\frac{16}{3}\right)^2 - 1\right] \equiv \zeta_{\rm cr.}
\end{equation}
and
\begin{equation}  \label{bound2}
 0> \beta\geq \beta_{\rm cr.} \equiv  -\frac{76}{153}\left[ \left(\frac{16}{3}\right)^2 - 1\right]^{-1} ~~.
\end{equation}

Given $b_a=0$, the EoM of the Lagrangian (\ref{model}) take the form 
\begin{equation} \label{EOM}
\left[f_{1}(X)+2f_{2}(X)R\right] G_{\mu\nu}+f_{2}(X)R^{2}g_{\mu\nu}
+\mathcal{D}_{\mu\nu}=-72f_{2}(X)\nabla_{\mu} \bar{X} \nabla_{\nu} X +g_{\mu\nu}\left[ V(\bar{X},X) +
144 f_{2}(X)\nabla^{\rho} \bar{X} \nabla_{\rho} X\right]~~,
\end{equation}
where we have introduced the notation
\begin{equation}\label{nota}
f_{1}(X)=\left(\frac{\beta}{2}-\frac{2}{M^{2}}|X|^{2}+\frac{11}{12}\zeta|X|^{4}\right)~,\quad 
f_{2}(X)=\left(\frac{1}{12M^{2}}-\frac{\zeta|X|^{2}}{72} \right)~, \quad G_{\mu\nu}=R_{\mu\nu}-\frac{1}{2}g_{\mu\nu}R~~,
\end{equation}
 and 
\begin{equation}\label{extraDer}
\mathcal{D}_{\mu\nu}(X,g_{\mu\nu})=2R(-\nabla_{\mu}\nabla_{\nu}+g_{\mu\nu}\Box )f_{2}(X)+2f_{2}(X)(-\nabla_{\mu}\nabla_{\nu}+g_{\mu\nu}\Box )R+(-\nabla_{\mu}\nabla_{\nu}+g_{\mu\nu}\Box )f_{1}(X) ~~,
\end{equation}
with $\Box=(1/\sqrt{-g})\partial_{\nu}\sqrt{-g}g^{\nu\mu}\partial_{\mu}$.
 
The EoM (\ref{EOM}) can be rewritten to the "Einstein-like" form 
\begin{equation} \label{EOM2}
G_{\mu\nu}=-\frac{1}{f_{1}(X)+2f_{2}(X)R }\left[ g_{\mu\nu}\left\{ f_{2}(X)R^{2}-
144f_{2}(X)\nabla^{\rho} \bar{X} \nabla_{\rho} X-V(X)\right\}+\tilde{\mathcal{D}}_{\mu\nu}\right]
\equiv T_{\mu\nu}^{G}~~,
\end{equation}
where we have introduced  the effective energy-momentum tensor  $T_{\mu\nu}^{G}$ with  
$\tilde{\mathcal{D}}_{\mu\nu}=\mathcal{D}_{\mu\nu}+72f_{2}(X)\nabla_{\mu} \bar{X} \nabla_{\nu} X$.

To further simplify our investigation, we assume slowly-varying fields by ignoring the $\tilde{\mathcal{D}}_{\mu\nu}$
term, and take the $X$-field in a Higgs phase with $X=\VEV{X}=X_0$ at $V'(X_0)=0$. It also implies 
\begin{equation} \label{3vev}
\VEV{f_1(X)} \equiv \bar{f}_1=\frac{1}{2}~,\quad \VEV{f_2(X)} \equiv \bar{f}_2=\frac{1}{12M^2_0} \quad{\rm and}
\quad \VEV{V(X)} =V(X_0) \equiv V_0~~.  
\end{equation}

In the FLRW background
\begin{equation} \label{flrw}
ds^{2}=-dt^{2}+a(t)^{2}[dr^{2}+r^{2}(d\theta^{2}+\sin^{2}\theta d\phi^{2})]
\end{equation}
the modified Friedman and Raychaudhuri equations without $X$-matter are given by 
\begin{equation} \label{fried}
\left(\frac{\dot{a}}{a}\right)^{2}-\frac{1}{6(\bar{f}_{1}+2\bar{f}_{2}R)}\left\{\bar{f}_{2}R^{2} +6\left(\frac{\dot{a}}{a}\right)\dot{R}\bar{f}_{2}-V_0
\right\}=0
\end{equation}
and
\begin{equation} \label{rai}
\left(\frac{\ddot{a}}{a}\right)+\frac{1}{2(\bar{f}_{1}+2\bar{f}_{2}R)}\left\{2\bar{f}_{2}\left(\frac{\dot{a}}{a}\right)\dot{R}+2\ddot{R}\bar{f}_{2}+\frac{1}{3}\bar{f}_{2}R^{2} -\frac{1}{3}V_0\right\}=0~~,
\end{equation}
respectively.
These equations can be rewritten to the standard form by introducing the effective gravitational pressure $p^{G}$
and density $\rho^{G}$ after decomposing the energy-momentum tensor as  
\begin{equation} \label{decomp}
T_{\mu\nu}^{G}=\rho^{G}u_{\mu}u_{\nu}+p^{G}(g_{\mu\nu}+u_{\mu}u_{\nu})~~.
\end{equation}
We find
\begin{equation} \label{den}
\rho_{G}=\frac{1}{2(\bar{f}_{1}+2\bar{f}_{2}R)}\left\{\bar{f}_{2}R^{2}+V_0+6\bar{f}_{2}\left(\frac{\dot{a}}{a}\right)\dot{R}\right\}
\end{equation}
and
\begin{equation} \label{pre}
p_{G}=-\frac{1}{2(\bar{f}_{1}+2\bar{f}_{2}R)}\left\{\bar{f}_{2}R^{2}+V_0
+4\bar{f}_{2}\left(\frac{\dot{a}}{a}\right)\dot{R}+2\bar{f}_{2}\ddot{R}\right\}~~.
\end{equation}

It implies  
\begin{equation} \label{prhoG}
\eqalign{
\rho_{G}+p_{G} & =  \frac{\bar{f}_{2}}{(\bar{f}_{1}+2\bar{f}_{2}R)}\left[-\ddot{R}+\left(\frac{\dot{a}}{a}\right)\dot{R}\right]
\cr
& = -\frac{f_{2}}{\bar{f}_{1}+2\bar{f}_{2}(6\dot{H}+12H^{2})}[6\ddot{H}+24\dot{H}^{2}+18H\ddot{H}-24H^{2}\dot{H}]
\cr }
\end{equation}
and
\begin{equation} \label{prhoG2} 
\rho_{G}+3p_{G}=-\frac{1}{2(\bar{f}_{1}+2\bar{f}_{2}R)}\left(2\bar{f}_{2}R^{2}+2V_0+6\bar{f}_{2}\left(\frac{\dot{a}}{a}\right)\dot{R}
+6\bar{f}_{2}\ddot{R}\right)~~.
\end{equation}

In the FLRW case, we have
\begin{equation}  \label{grav3}
R=6\left(\frac{\ddot{a}}{a}+\frac{\dot{a}^{2}}{a^{2}} \right)=6\dot{H}+12H^{2}~,\quad
\dot{R}=6\ddot{H}+24H\dot{H} \quad {\rm and} \quad
\ddot{R}=6\dddot{H}+24\dot{H}^{2}+24H\ddot{H}~~.
\end{equation}

The physically relevant situations (inflation, dark energy, black holes) correspond to $R\geq 0$, so that the condition 
$\bar{f}_{1}+2\bar{f}_{2}R>0$ or, equivalently, $R> - 3M_0^2$, always applies.  Equations  (\ref{den}),  (\ref{prhoG}) and  (\ref{prhoG2}) are needed for studying  the energy conditions in the next Section.

\section{The energy conditions}

\vglue.2in
The {\bf NEC} reads
\begin{equation} \label{nec1}
T_{\mu\nu}n^{\mu}n^{\nu}\geq 0~~,
\end{equation}
where $n^{\mu}$ is a generic null-like vector. In the  FLRW background it corresponds to
\begin{equation} \label{nec2}
\rho^{G}+p^{G}\geq 0~~.
\end{equation}
Hence, in our case (\ref{prhoG}), we have
\begin{equation} \label{nec3}
0\leq p^{G}+\rho^{G}=\frac{\bar{f}_{2}}{\bar{f}_{1}+2\bar{f}_{2}R}\left[-\ddot{R}+H\dot{R}\right]~~.
\end{equation}
When $\dot{R}=0$, this condition is satisfied in its extremal bound corresponding to $w=-1$. The NEC is generically satisfied 
when  $\ddot{R}<<H\dot{R}$.  For instance, given the inflationary conditions $H\gg M_0$ and $\dot{H}\simeq 0$, the NEC is always satisfied. 
\vglue.2in

The {\bf WEC} reads
\begin{equation} \label{wec1}
T_{\mu\nu}u^{\mu}u^{\nu}\geq 0~~,
\end{equation}
where $u^{\mu}$ is a generic time-like vector.  The corresponding FLRW condition reads
\begin{equation} \label{wec2}
\rho^{G}\geq 0
\end{equation}
and, in our case (\ref{den}), takes the form
\begin{equation} \label{wec3}
\frac{1}{\bar{f}_{1}+2\bar{f}_{2}R}\left[ \bar{f}_{2}R^{2}+V_0 + 6\bar{f}_{2}\left(\frac{\dot{a}}{a}\right)\dot{R}\right] \geq 0~~.
\end{equation}
In the case of slowly varying fields, it amounts to 
\begin{equation} \label{wec4}
\frac{R^2}{12M_0^2} + V_0 \geq 0~~. 
\end{equation}
In a generic case, (\ref{wec3}) is easily satisfied as long as $R,\dot{R}\geq 0$ and $V_0\geq 0$. The restrictions on the parameters
for  $V_0\geq 0$ are given in the previous Section.
\vglue.2in

The {\bf SEC} reads
\begin{equation} \label{sec1}
\left(T_{\mu\nu}-\frac{1}{2}Tg_{\mu\nu}\right)u^{\mu}u^{\nu}\geq 0~~,
\end{equation}
where $u^{\mu}$ is a generic time-like vector. In the  FLRW case it reads
\begin{equation} \label{sec2}
\rho+3p^{G}\geq 0
\end{equation}
and because of (\ref{prhoG2}) takes the form
\begin{equation} \label{sec3}
-\frac{1}{(\bar{f}_{1}+2\bar{f}_{2}R)}\left[ \bar{f}_{2}R^{2}+V_0+ 3\bar{f}_{2}\left(\frac{\dot{a}}{a}\right)\dot{R}+
3\bar{f}_{2}\ddot{R}\right]\geq 0~~.
\end{equation}
This condition in the case of slowly varying fields amounts to 
\begin{equation} \label{sec4}
\frac{R^2}{12M_0^2} + V_0 \leq 0
\end{equation}
that is the opposite to (\ref{wec4}). The condition (\ref{sec3}) can never be satisfied in an expanding universe when
 $R,\dot{R},\ddot{R},\dot{a}/a>0$ and $V_0\geq 0$.

\section{Conclusion and outlook}

In this paper we applied the common energy conditions to the Starobinsky supergravity theories (in the old-minimal supergravity setup). We focused on the simplest model with a stable vacuum. Since the axion field is stabilized at its vanishing value, we ignored its dynamics together
with all fermions, and restricted ourselves to the $(X,R)$-sector of the model. Even this simplified model is highly non-trivial with the complicated Lagrangian (\ref{model}) due to the presence of the non-minimal couplings of the complex scalar field $X$ to both $R$ and $R^2$ terms in
the gravitational sector and the non-minimal self-interactions of the scalar field itself. Nevertheless, its scalar potential is fixed up to a few parameters. The cosmological constant $V_0$ can be non-negative though it leads to the tight constraints on the parameters of the model. When the cosmological constant is positive, supersymmetry is spontaneously broken, while R-symmetry can be preserved. 

We calculated the EoM, identified the supergravity contribution to the effective energy-momentum tensor, and then applied our results to the three energy conditions: NEC, WEC and SEC. These conditions take the form of inequalities for 
the certain complicated quantities evolving in time when considering the complex scalar field $X$ as a dynamical field. In the case 
when the scalar field $X$ is sitting in the minimum of its scalar potential (the Higgs phase),  those constraints are greatly simplified. 
In such case, the WEC can be easily satisfied, when back reaction is ignored. The NEC is saturated to $w=-1$. It appears to be impossible to 
satisfy both WEC and SEC  simultaneously. The latter may not be surprising because the SEC violation is a generic feature of inflation and
dark energy.

Though the Starobinsky supergravity appears to be compatible with both NEC and WEC after fine tuning of the parameters, it apparently requires large positive values of $\zeta$ that seems to be rather unnatural from the phenomenological viewpoint. Our conclusions are the direct consequence of the dynamical nature of the $X$-field that has to be stabilized by the $\zeta$-term.

Instead of adding the stabilization term to the simplest  $N$-function given by the first term in Eq.~(\ref{ansatz}), one can impose the nilpotency condition,
 $\mathcal{R}^2=0$,  on the chiral curvature superfield $\mathcal{R}$ along the lines of Refs.~\cite{Rocek:1978nb,Komargodski:2009rz} and thus get rid of the
$X$-field. It leads to the so-called {\it Volkov-Akulov-Starobinsky supergravity} \cite{Antoniadis:2014oya} with the sinflaton originating from 
$b_a$ as the only extra bosonic d.o.f. In this theory one may eliminate violations of NEC and WEC, apart from concern about possible loss of unitarity.

\begin{acknowledgments} 

A.A. is partially supported  by the MIUR research grant "Theoretical Astroparticle Physics" PRIN 2012CPPYP7.
 S.V.K. is supported by a Grant-in-Aid of the Japanese Society for Promotion of Science (JSPS) under No.~26400252, a TMU President Grant of Tokyo Metropolitan University in Japan, the World Premier International Research Center Initiative (WPI Initiative), MEXT, Japan, and the Competitiveness Enhancement Program of Tomsk Polytechnic University in Russia.

\end{acknowledgments}

\end{document}